\documentclass[letterpaper,twocolumn,10pt]{article}
\usepackage{usenix2025_SOUPS}

\usepackage{filecontents}
\usepackage{graphicx}
\usepackage[most]{tcolorbox}
\usepackage{xcolor}
\usepackage{tikz}
\usetikzlibrary{shapes.geometric, arrows, positioning}

\newtcolorbox{prompt}[1][]{
    colframe=black,
    colback=white!95!gray,
    fonttitle=\bfseries,
    title=#1,
    before upper={\obeylines},
    breakable,
    sharp corners
}

\newcommand{\attackSession}[5]{
    \begin{prompt}[#1]

    \textbf{Attack Strategy:}  
    #2

    \vspace{5pt} \hrule \vspace{5pt}

    \textbf{Attack Message:}  
    \textcolor{red!80!black}{#3}

    \vspace{5pt} \hrule \vspace{5pt}

    \textbf{Victim's Thought:}  
    \textcolor{blue!80!black}{#4}

    \vspace{5pt}

    \textbf{Visit Likelihood:}  
    \textcolor{black!80!black}{\Large \textbf{#5}}

    \end{prompt}
}

\definecolor{myBlue}{HTML}{0072B2}
\definecolor{myOrange}{HTML}{E69F00}
\definecolor{myPurple}{HTML}{CC79A7}

\tikzstyle{startstop} = [rectangle, rounded corners, minimum width=5.5cm,
    minimum height=0.7cm, text centered, align=center, draw=black, fill=myBlue!30]
\tikzstyle{process}   = [rectangle, minimum width=5.5cm, minimum height=0.7cm,
    text centered, align=center, draw=black, fill=myOrange!30]
\tikzstyle{decision}  = [diamond, aspect=4, text centered, align=center, draw=black, fill=myPurple!30]
\tikzstyle{arrow}     = [thick,->,>=stealth]

\begin{document}

\date{}

\title{\Large \bf Can We End the Cat-and-Mouse Game? Simulating Self-Evolving Phishing Attacks with LLMs and Genetic Algorithms}

\def\plainauthor{}

\author{
{\rm Seiji Sato}\\
Shizuoka University\\
sato.seiji.a@shizuoka.ac.jp
\and
{\rm Tetsushi Ohki}\\
Shizuoka University, RIKEN AIP\\
ohki@inf.shizuoka.ac.jp
\and
{\rm Masakatsu Nishigaki}\\
Shizuoka University\\
nisigaki@inf.shizuoka.ac.jp
} 

\maketitle

\begin{abstract}

Anticipating emerging attack methodologies is crucial for proactive cybersecurity. Recent advances in Large Language Models (LLMs) have enabled the automated generation of phishing messages and accelerated research into potential attack techniques. However, predicting future threats remains challenging due to reliance on existing training data. To address this limitation, we propose a novel framework that integrates LLM-based phishing attack simulations with a genetic algorithm in a psychological context, enabling phishing strategies to evolve dynamically through adversarial interactions with simulated victims. Through simulations using Llama 3.1, we demonstrate that (1) self-evolving phishing strategies employ increasingly sophisticated psychological manipulation techniques, surpassing naive LLM-generated attacks, (2) variations in a victim's prior knowledge significantly influence the evolution of attack strategies, and (3) adversarial interactions between evolving attacks and adaptive defenses create a cat-and-mouse dynamic, revealing an inherent asymmetry in cybersecurity—attackers continuously refine their methods, whereas defenders struggle to comprehensively counter all evolving threats. Our approach provides a scalable, cost-effective method for analyzing the evolution of phishing strategies and defenses, offering insights into future social engineering threats and underscoring the necessity of proactive cybersecurity measures.

\end{abstract}

\section{Introduction}\label{sec:introduction}

Cybersecurity is often described as a continuous cat-and-mouse game, where defenders must constantly adapt to evolving attack strategies. To effectively safeguard systems and sensitive information against emerging threats, it is crucial to anticipate and understand future attack methodologies. Prior research has addressed this challenge by analyzing potential attack scenarios; however, traditional social engineering studies have relied on costly human experiments to assess attack effectiveness and analyze victim response behavior. Such approaches inherently limit the exploration of novel attack methods under varying conditions.

One potential solution is to use simulations for threat analysis. Simulations have been widely employed across domains such as computer systems, network infrastructures, and cyber-physical systems \cite{kavak2021simulation, jaber2022towards, yohanandhan2020cyber}. Despite their effectiveness, modeling social engineering attacks presents unique challenges due to the complexity of human behavior. While some studies have attempted to model human decision-making using mathematical frameworks \cite{kavak2021simulation}, human psychology and behavior are inherently complex and cannot be easily represented by such models, limiting these approaches’ capacity to accurately capture human responses.

Recent advances in large language models (LLMs) have demonstrated their potential in generating automated attacks. Studies have shown that LLMs can generate realistic phishing messages \cite{hazell2023spear, charfeddine2024chatgpt, marchal2024generative, heiding2024devising}. Moreover, research has explored multi-agent approaches using LLMs to conduct sophisticated and adaptive attacks for computer system threat analysis \cite{deng2024pentestgpt, xu2024autoattacker, fang2024teams}. However, these approaches rely on the existing knowledge embedded in LLM training data, making them unsuitable for predicting emerging threats, and they tend to focus on simulating the attacker while leaving the modeling of the defender’s human behavior as an open challenge.

Since the effectiveness of an attack depends on various victim-specific factors, incorporating these variables into threat analysis is crucial. In particular, understanding the impact of phishing awareness and knowledge is essential, as these factors can be improved through education, public awareness campaigns, and security guidelines. However, much of the existing research overlooks variations in victim knowledge and instead focuses on generating universally effective attacks. While some studies have attempted to personalize phishing messages by providing victim-specific information to LLMs \cite{hazell2023spear, heiding2024devising} or by modeling victims using LLMs to simulate and evaluate phishing effectiveness \cite{asfour2023harnessing}, these approaches primarily validate known phishing strategies rather than analyzing the emergence of new threats shaped by attacker-defender interactions. Similarly, although LLMs have been explored for generating security guidelines \cite{mcintosh2023harnessing, quinn2024applying} and detecting phishing attempts \cite{koide2024chatphishdetector, nahmias2024prompted}, these studies concentrate on countering known attacks rather than anticipating future attack vectors.

To overcome these limitations, we propose a novel framework that integrates LLM-based simulations of attacker-victim interactions with a genetic algorithm to enable the self-evolution of phishing strategies. Instead of generating attacks based on a single inference of an LLM, we employ evolutionary methods that iteratively refine attack strategies based on their effectiveness against simulated victims. This approach enables the generation of novel attacks that are not constrained by the static knowledge embedded in LLM training data.

In our framework, the attacker’s strategy is represented as a natural language prompt that guides the LLM in generating phishing messages rather than directly producing the messages themselves. This representation allows us to analyze the psychological mechanisms underlying phishing strategies rather than merely their surface-level characteristics. By doing so, we gain deeper insights into the psychological techniques that contribute to phishing effectiveness, many of which have been identified as critical factors in prior research \cite{longtchi2024internet, yasin2019contemplating, mouton2014social, montanez2020human}. Additionally, we incorporate psychological principles into the mutation process of our genetic algorithm, including those not previously linked to phishing, to uncover effective attack techniques beyond those already documented in existing literature.

Additionally, we simulate the victim’s response using an LLM to better capture the complexity of human decision-making. Previous psychological research has demonstrated that LLMs can model human behavior in social contexts \cite{dillion2023can, mittelstadt2024large, ke2024exploring}. Leveraging this capability, we design the LLM to act as a phishing victim that evaluates messages and assigns a score to each attack. This score then guides the genetic algorithm in refining phishing strategies. To account for variations in defender behavior, we incorporate different levels of phishing awareness into the LLM’s input prompt, allowing us to analyze how attack strategies adapt to victims with varying knowledge levels. Furthermore, we model the dynamic co-evolution of attackers and defenders by iteratively updating the victim’s knowledge based on the phishing messages encountered during the evolutionary process. This mechanism mirrors the adversarial learning dynamics of Generative Adversarial Networks (GANs), enabling us to study the interactive evolution of attack strategies and defense strategies.

Our approach offers a scalable and cost-effective method for analyzing the dynamic interplay between evolving attack and defense strategies while preserving the complexity of human factors. By leveraging LLM-driven simulations and evolutionary algorithms, we uncover insights into the emergence of novel phishing strategies, adaptive adversarial behaviors, and the development of proactive countermeasures. Ultimately, this framework not only enhances our understanding of phishing evolution but also provides a foundation for designing more effective security measures against future threats.

\paragraph{Research Questions.} To systematically explore the impact of our proposed framework, we address the following research questions:

\newcommand{\rqOne}{RQ1: How does the integration of genetic algorithms with LLM-based simulations, incorporating underlying psychological mechanisms, enhance the sophistication and effectiveness of phishing attack strategies over iterative evolutions?}
\newcommand{\rqTwo}{RQ2: How do variations in a victim’s prior phishing awareness influence the adaptation and evolution of attack strategies in an iterative simulation?}
\newcommand{\rqThree}{RQ3: How does the iterative adaptation of both phishing attack strategies and victim defenses shape the evolution of adversarial interactions over time?}

\begin{itemize}
    \item \rqOne
    \item \rqTwo
    \item \rqThree
\end{itemize}

\paragraph{Contributions.} In addressing these research questions, this paper makes the following key contributions:

\begin{itemize}
    \item We propose a novel framework that enables the iterative evolution of phishing attack strategies through LLM-driven simulations and genetic algorithms, incorporating underlying psychological mechanisms to enhance sophistication and effectiveness.
    \item We systematically analyze how phishing strategies evolve in response to variations in victim awareness, highlighting potential security risks posed by adaptive attacks that can exploit gaps in defensive strategies.
    \item We examine the co-evolutionary dynamics between phishing attack strategies and defensive mechanisms, revealing adaptive patterns that inform the development of proactive cybersecurity strategies.
\end{itemize}

\section{Related Work}

\subsection{Simulation in Social Engineering}

Traditionally, social engineering research has relied on human-based experiments to analyze responses to attacks. Several survey papers summarize these efforts, such as \cite{longtchi2024internet, yasin2019contemplating, syafitri2022social}. While these studies provide valuable insights into real human behavior, they are costly, time-consuming, and limited in their ability to explore diverse attack scenarios at scale.

To address these challenges, simulation-based approaches have been introduced. Kavak et al. \cite{kavak2021simulation} provide a comprehensive survey of simulation techniques in cybersecurity, including mathematical models used to simulate human decision-making. However, these models struggle to capture the complexity of human psychology and behavior, limiting their applicability in analyzing social engineering threats where human factors play a crucial role.

The emergence of LLMs has significantly advanced phishing attack research by enabling the automated generation of natural and sophisticated phishing messages. Several studies have leveraged LLMs to generate phishing content and assess its effectiveness \cite{hazell2023spear, charfeddine2024chatgpt, marchal2024generative, heiding2024devising}. While these methods automate phishing attack creation, most still require human participants to evaluate the attack’s effectiveness. Some research has explored simulating victim behavior using LLMs to mimic human responses. For instance, Asfour et al. \cite{asfour2023harnessing} introduced a method that incorporates Big Five personality traits into GPT-4 prompts to simulate human reactions to phishing emails. However, this approach does not account for the dynamic evolution of attack strategies, limiting its usefulness in predicting future threats.

Beyond cybersecurity, psychological research has employed LLMs to simulate human cognitive and behavioral patterns \cite{huang2023humanity, ke2024exploring, tshimula2024psychological}. These studies demonstrate the potential of LLM-driven simulations for large-scale psychological experiments, though they have not been explicitly applied to cybersecurity and social engineering threats.

\subsection{Autonomous Attack Generation Using LLMs}

Prior to LLM advancements, several techniques were proposed for automating phishing text generation and social engineering attacks \cite{huber2009towards, lauinger2010honeybot, seymour2016weaponizing}. These approaches utilized machine learning, natural language processing, and rule-based algorithms. However, they were inherently constrained by existing datasets and predefined logic, preventing them from generating novel attack strategies.

With the integration of LLMs and autonomous agent frameworks, more sophisticated and contextually adaptive attack methodologies have emerged. Deng et al. \cite{deng2024pentestgpt} and Xu et al. \cite{xu2024autoattacker} have explored LLM-based penetration testing frameworks where multiple LLM agents perform tasks autonomously under a central planning agent. Fang et al. \cite{fang2024teams} demonstrated how LLM-driven sub-agents can exploit zero-day vulnerabilities through exploratory interactions. However, these methods primarily target traditional computer systems and do not focus on social engineering or human-specific attack strategies.

Other research efforts have investigated generating new phishing messages from existing datasets. Afane et al. \cite{afane2024next} proposed prompting LLMs to transform existing phishing emails, while Chen et al. \cite{chen2024adapting} used a framework integrating LLMs with Generative Adversarial Networks (GANs) to synthesize new phishing content. However, these approaches primarily produce variations of existing phishing messages rather than creating novel, situationally adaptive attacks.

Additionally, recent work has explored optimizing LLM-generated inputs—including prompts and adversarial attack strategies—using evolutionary algorithms and fuzzing techniques \cite{guo2023connecting, brownlee2023enhancing, braberman2024tasks, yu2024promptfuzz, yu2024llm}. While these methods facilitate the discovery of previously unknown inputs, they have not been specifically applied to the evolution of phishing attack strategies that adapt to varying victim conditions.

\subsection{Generating Countermeasures Using LLMs}

Research on LLM-driven countermeasures has primarily focused on phishing detection and the generation of defensive strategies. Unlike black-box detection techniques, LLM-based approaches provide interpretability, enabling the creation of human-readable guidelines and policies. Studies such as \cite{mcintosh2023harnessing, quinn2024applying} have explored using LLMs to generate security policies, though these efforts primarily address known threats and do not consider the dynamic evolution of phishing strategies.

Phishing detection research has also leveraged LLMs to enhance interpretability. Koide et al. \cite{koide2024chatphishdetector} proposed an LLM-based phishing detection model that provides human-readable justifications for its decisions. Similarly, Nahmias et al. \cite{nahmias2024prompted} explored using prompt-based phishing detection models that allow users to input and modify detection criteria. While these methods help in understanding phishing characteristics, they rely on known attack patterns embedded in LLM training data and do not account for evolving attack strategies.

Overall, while significant progress has been made in LLM-based simulation, automated attack generation, and defense mechanisms, existing approaches largely focus on static or predefined attack strategies. Our proposed framework addresses this limitation by integrating LLM-driven phishing simulations with a genetic algorithm, enabling attacks to evolve autonomously based on victim interactions and conditions. This adaptive approach enhances the realism of simulations and provides a more effective means of analyzing and anticipating emerging phishing threats.

\section{Methodology}\label{sec:method}

\subsection{Attack Settings}

We model an attacker and a victim as actors in a simulation, interacting on a social media platform. The attacker crafts a targeted phishing message and sends it as a mention message to the victim. This message contains a URL leading to a website, with the goal of inducing the victim to visit it. While we do not simulate the website itself or the victim's behavior post-click, we assume the site is a phishing page designed to extract sensitive credentials.

Although various communication channels such as email and SMS can be used for phishing, we opt for a social media platform due to its ability to minimize situational dependencies and external influencing factors. Common phishing techniques, such as Deceptive Phishing and Spear Phishing via email and SMS \cite{chanti2022literature}, present additional complexities. Deceptive Phishing, which impersonates institutions like banks, often triggers safety mechanisms in large language models (LLMs), necessitating jailbreak techniques. Spear Phishing, highly dependent on contextual settings, introduces numerous variables that complicate controlled simulations. Conversely, social media mentions allow for a more straightforward simulation where phishing messages can be naturally generated without bypassing LLM restrictions, while requiring minimal situational setup.

\subsection{Overall Procedure}\label{sec:method-overall-procedure}

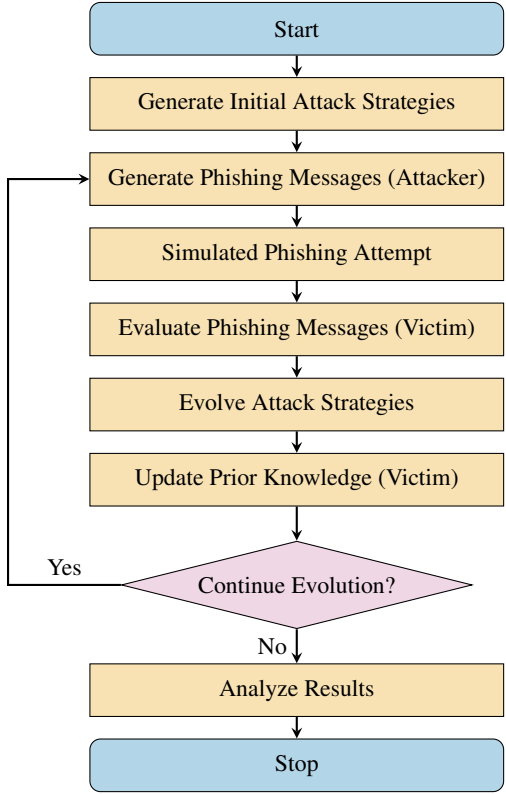
\begin{figure}[tb]
    \centering
    \begin{tikzpicture}[node distance=1cm, auto, font=\small]
      \node (start)   [startstop] {Start};
      \node (init)    [process, below of=start] {Generate Initial Attack Strategies};
      \node (attacker)[process, below of=init] {Generate Phishing Messages (Attacker)};
      \node (sim)     [process, below of=attacker] {Simulated Phishing Attempt};
      \node (evaluate)[process, below of=sim] {Evaluate Phishing Messages (Victim)};
      \node (ga)      [process, below of=evaluate] {Evolve Attack Strategies};
      \node (update)  [process, below of=ga] {Update Prior Knowledge (Victim)};
      \node (decide)  [decision, below of=update, yshift=-0.4cm] {Continue Evolution?};
      \node (analyze)[process, below of=decide, yshift=-0.4cm] {Analyze Results};
      \node (stop)    [startstop, below of=analyze] {Stop};
    
      \draw [arrow] (start) -- (init);
      \draw [arrow] (init) -- (attacker);
      \draw [arrow] (attacker) -- (sim);
      \draw [arrow] (sim) -- (evaluate);
      \draw [arrow] (evaluate) -- (ga);
      \draw [arrow] (ga) -- (update);
      \draw [arrow] (update) -- (decide);
      \draw [arrow] (decide.west) -- node[anchor=south] {Yes} ++(-1.5,0) |- (attacker.west);
      \draw [arrow] (decide) -- node[anchor=east] {No} (analyze);
      \draw [arrow] (analyze) -- (stop);
    \end{tikzpicture}
    \caption{Overall procedure of our methodology}
    \label{fig:method-overall-procedure}
\end{figure}

Figure~\ref{fig:method-overall-procedure} outlines the overall procedure of our methodology. Due to the length of the detailed prompts given to the LLM in the following procedure, they are included in Appendix~\ref{sec:app-method-prompts}.

\paragraph{1. Generation of Initial Attack Strategies}
We utilize an LLM to generate an initial set of phishing attack strategies based solely on its pre-trained knowledge. Since we employ genetic algorithms, we generate a predefined number of strategies, determined by the scale of the simulation.

As discussed in Section~\ref{sec:introduction}, we represent an attack strategy as a natural language prompt that guides the LLM in generating phishing messages. The use of natural language allows for a high degree of flexibility in representing complex attack techniques while maintaining human interpretability, which facilitates the extraction of insights for defensive measures.

\paragraph{2. Generation of Phishing Messages (Attacker)}
For each attack strategy, we prompt the LLM to generate corresponding phishing messages. The strategy serves as an input to the LLM, guiding it to produce phishing messages in the role of an attacker. Given the inherent randomness of LLM outputs, we generate multiple phishing messages per strategy to mitigate variance in effectiveness.

\paragraph{3. Simulated Phishing Attempt}
Each generated phishing message is sent to a simulated victim, representing a phishing attempt within the framework.

\paragraph{4. Evaluation of Phishing Messages (Victim)}
Upon receiving a phishing message, the victim evaluates its likelihood of inducing a URL click. We prompt the LLM, acting as the victim, to rate the likelihood of clicking on the link using a scale from 1 (Not likely at all) to 10 (Very likely). To account for variations in user knowledge, we incorporate prior knowledge into the prompt, allowing the model to contextualize the message and assess its credibility accordingly.

To ensure a more reasoned evaluation, we employ the Chain-of-Thought (CoT) reasoning approach \cite{wei2022chain}, where the LLM first generates a thought process before providing a final likelihood score. Each message is evaluated independently to prevent inter-message influence.

\paragraph{5. Evolution of Attack Strategies Using Genetic Algorithms}
For each attack strategy, we compute an effectiveness score based on the average visit likelihood of the phishing messages it generates. This score serves as the fitness value in our genetic algorithm. If the likelihood values from 1 to 10 are used directly, the impact of the difference in likelihood on the selection probability diminishes as the likelihood increases. To counteract this effect and ensure that higher-performing strategies maintain a consistent advantage, we define the fitness score as $1.4^v$, where $v$ is the average visit likelihood. This exponential scaling amplifies differences in selection probability across varying likelihood values, ensuring a more pronounced distinction between strategies.

Once fitness scores are computed, they are passed to the genetic algorithm module, which generates new attack strategies through evolutionary processes. A detailed description of the genetic algorithm mechanism is provided in Section~\ref{sec:ga-operations}.

\paragraph{6. Updating the Victim's Prior Knowledge (Victim, Co-Evolution Simulation)}
In scenarios where we simulate co-evolution between the attacker and the defender, the victim’s prior knowledge is periodically updated. Specifically, we sample the top 10 phishing messages with the highest visit likelihood scores and input them into the LLM along with the current prior knowledge. The choice of 10 messages is deliberate—limiting the input prevents the LLM from being overwhelmed by excessive information, ensuring a more consistent and reliable refinement of defensive heuristics. This enables the LLM to refine its defensive heuristics, dynamically adapting to evolving phishing strategies.

\paragraph{7. Continuous Evolution}
The process iterates from Step 2 (Generation of Phishing Messages) to Step 6 (Updating the Victim’s Prior Knowledge), ensuring a continuous evolutionary cycle where attack strategies evolve and defensive knowledge adapts over successive generations.

\paragraph{8. Analysis of Results}
We analyze the attack strategies, victim responses, and refined prior knowledge to extract key insights into the evolution of phishing strategies and potential countermeasures.

\subsection{Genetic Algorithm Operations}\label{sec:ga-operations}

This section details the genetic algorithm operations used to evolve phishing attack strategies. As described in the previous section, the effectiveness of each strategy is evaluated based on the average visit likelihood assigned by the victim model. The fitness values are calculated from these effectiveness scores, which determine the selection of parent strategies for genetic operations.

\paragraph{Copy (Elitism)}
To maintain stability in the evolutionary process, a subset of the highest-scoring attack strategies is directly copied to the next generation without modification. This elitism mechanism ensures that the most effective strategies persist across generations, preventing the genetic algorithm from prematurely discarding high-performing solutions.

Following the Copy operation, the remaining strategies undergo either Crossover or Mutation to introduce variation. The number of copied strategies and the ratio of Crossover and Mutation are tunable parameters of the simulation. To select parent strategies for these operations, we employ a roulette wheel selection mechanism, which assigns selection probabilities proportional to fitness scores. This ensures that higher-performing strategies are more likely to be chosen while maintaining diversity in the evolving population.

\paragraph{Crossover}
In the Crossover operation, two selected parent strategies are combined to generate a new attack strategy. To achieve this, we provide both parent strategies as input to the LLM and prompt it to extract key elements from each. The LLM then synthesizes a new phishing strategy that integrates these elements in a novel manner. This operation facilitates the recombination of effective components from different strategies, potentially yielding superior attack techniques.

\paragraph{Mutation}
Mutation introduces targeted modifications to existing strategies, fostering exploration of new attack techniques. The mutation process consists of the following steps:

\begin{enumerate}
    \item A psychological theory name is randomly selected from a predefined list of 250 theories extracted from Wikipedia sources, including “List of social psychology theories” \cite{enwiki:1202707650}, “List of psychological effects” \cite{enwiki:1254715116}, and “List of cognitive biases” \cite{enwiki:1272835005}. This set is not exhaustive but encompasses a diverse selection of psychological principles relevant to persuasion and deception, providing sufficient variety for the mutation process.
    \item The LLM is prompted to generate a short description of the selected theory.
    \item The generated description and an existing attack strategy are provided as input to the LLM, which is then prompted to adapt the strategy by incorporating the psychological theory.
\end{enumerate}

This mutation process encourages the discovery of novel phishing techniques, including those based on psychological principles that have not been previously identified as effective in phishing. Some example theories include “Attribution Theory,” “Barnum Effect,” and “Cognitive Dissonance,” each of which provides unique insights into manipulating user behavior.

\begin{figure}[tb]
    \centering
    \begin{tikzpicture}[node distance=1cm, auto, font=\small]
      \node (start) [startstop] {Start GA Operations};
      \node (fitness) [process, below of=start] {Compute Fitness Scores};
      \node (copy) [process, below of=fitness] {Copy Elite Strategies};
      \node (select) [process, below of=copy] {Select Parents (Roulette Wheel)};
      \node (genetic) [process, below of=select] {Apply Genetic Operators\\ (Crossover/Mutation)};
      \node (decision) [decision, below of=genetic, yshift=-0.4cm] {Reached Population?};
      \node (formGen) [process, below of=decision, yshift=-0.4cm] {Form New Generation};
      \node (endGA) [startstop, below of=formGen] {End GA Operations};

      \draw [arrow] (start) -- (fitness);
      \draw [arrow] (fitness) -- (copy);
      \draw [arrow] (copy) -- (select);
      \draw [arrow] (select) -- (genetic);
      \draw [arrow] (genetic) -- (decision);
      \draw [arrow] (decision) -- node[anchor=east] {Yes} (formGen);
      \draw [arrow] (formGen) -- (endGA);
      \draw [arrow] (decision.west) -- node[anchor=south] {No} ++(-1.5,0) |- (select.west);
    \end{tikzpicture}
    \caption{Genetic algorithm procedure}
    \label{fig:method-genetic-algorithms}
\end{figure}
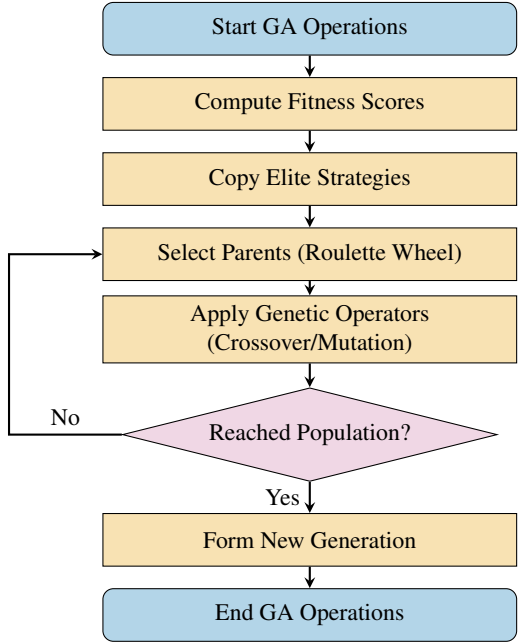

These genetic operations continue until the next generation reaches the predefined population size. The complete genetic algorithm procedure is illustrated in Figure~\ref{fig:method-genetic-algorithms}.

\section{Evaluation}\label{sec:evaluation}

This section presents the evaluation of our approach, guided by the following research questions:

\begin{itemize}
    \item \rqOne
    \item \rqTwo
    \item \rqThree
\end{itemize}

\subsection{Evaluation Settings}

Our evaluation was conducted under the following settings:

\noindent\textbf{Language Model:} LLaMA 3.1 8B \\
\noindent\textbf{Epochs:} 30 \\
\noindent\textbf{Attack Strategies per Generation:} 15 \\
\noindent\textbf{Phishing Messages per Strategy:} 3 \\
\noindent\textbf{Copied Strategies:} 3 \\
\noindent\textbf{Crossover Strategies:} 9 \\
\noindent\textbf{Mutated Strategies:} 3

Due to the large number of queries required for our approach, we employed LLaMA 3.1 8B, an open-source model that can be executed locally, thereby reducing operational costs. Since our method is model-agnostic, we chose LLaMA 3.1 8B as a representative LLM for evaluation. The model was hosted using Ollama \cite{ollama}, and the proposed system was implemented in Python, utilizing Mirascope \cite{mirascope} as a wrapper library for LLM interactions.

As a phishing URL, we used an imaginary Amazon product page (https://amazon.com/dp/123456) because the platform is widely known and naturally hosts login forms, making it a plausible phishing vector.

\subsection{Evaluation (RQ1)}\label{sec:evaluation-rq1}

We begin by analyzing the evolution of attack strategies in scenarios where the victim has no prior knowledge. To simulate this condition, we set the prior knowledge parameter to "N/A" in the prompt, which the victim uses to evaluate the likelihood of visiting a website based on the attacker's message.

\begin{figure}[tb]
\centering
\includegraphics[width=0.5\textwidth, trim=10 0 10 20, clip]{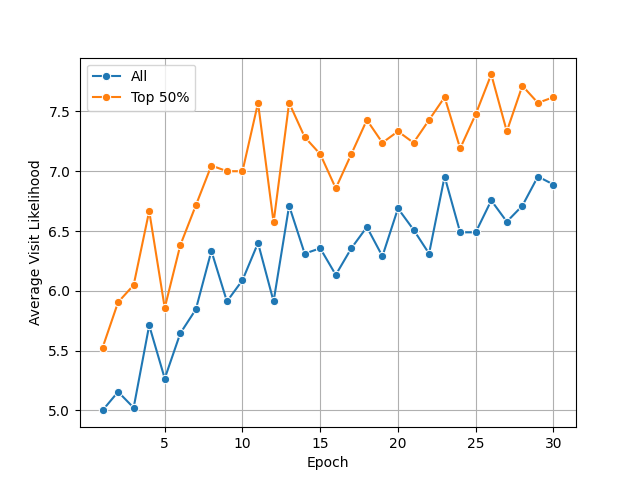}
\caption{Evolution of average visit likelihood (without victim's prior knowledge)}
\label{fig:eval-none-likelihood}
\end{figure}

Figure~\ref{fig:eval-none-likelihood} presents the average visit likelihood across all strategies, as well as the top 50\% of strategies with the highest likelihood. The results indicate a steady increase in visit likelihood as the number of epochs progresses, suggesting that the attack strategies become increasingly effective through evolutionary refinement.

\begin{figure}[tb]
\centering
\includegraphics[width=0.5\textwidth, trim=40 20 60 60, clip]{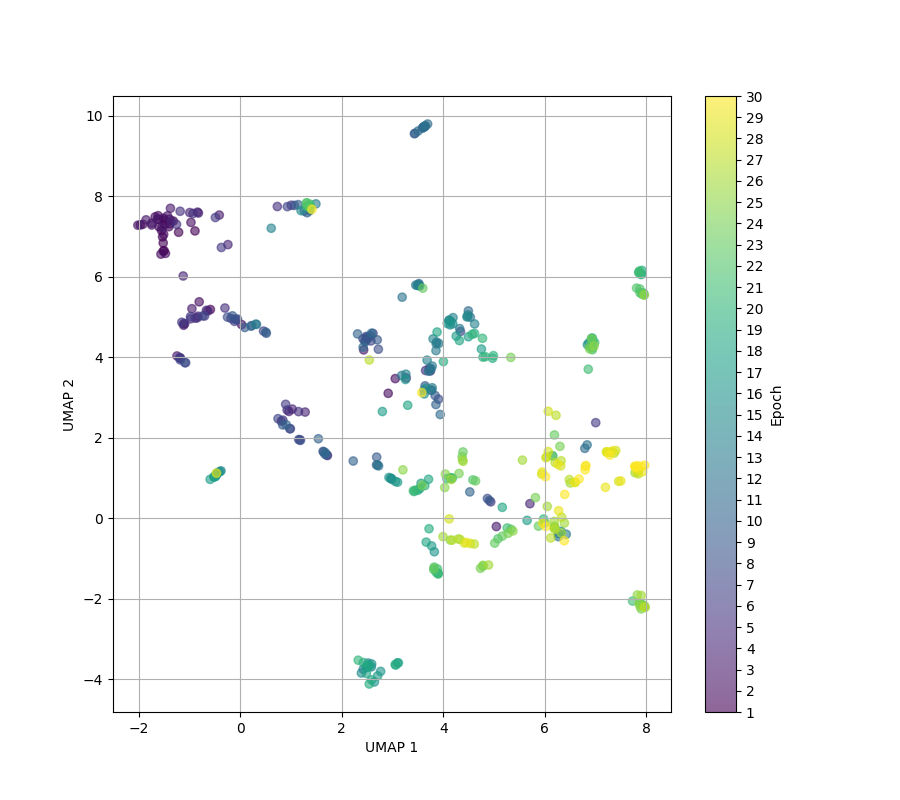}
\caption{Embeddings of attack strategies (without victim's prior knowledge)}
\label{fig:eval-none-embeddings}
\end{figure}

Figure~\ref{fig:eval-none-embeddings} visualizes the embeddings of attack strategies projected into two dimensions using UMAP. The results reveal that the embedding vectors, representing the semantic structure of the strategies, shift over epochs as the attack explores different directions.

To further understand the progression of attack strategies, we summarize below the key principles of the top 50\% strategies at epochs 1 and 30. These summaries are generated using the LLM, which extracts key themes from the strategies and distills them into a set of principles. The exact prompt used for this process is provided in Appendix~\ref{sec:app-evaluation-rq1}. The identified principles highlight the transition from direct strategies—such as exploiting curiosity and urgency—to more refined and persuasive techniques. At epoch 1, strategies primarily emphasize strong incentives and urgency, whereas by epoch 30, they evolve to incorporate more nuanced persuasion, such as aligning with user values and subtly reinforcing their expertise.

\begin{prompt}[Epoch 1 Principles]
**Principle 1: Spark Curiosity**
Craft mention messages that pique interest by highlighting exclusive benefits or value offered on the website, using a conversational tone, and creating a sense of urgency through limited-time offers or scarcity.

**Principle 2: Personalize and Add Urgency**
Make mention messages relevant and actionable by personalizing them to the user's interests or needs, while also showcasing a clear call-to-action and emphasizing limited-time access to valuable resources or exclusive offers.

**Principle 3: Show Personality and Unique Value**
Use conversational tone and concise language in mention messages to showcase your brand's personality and highlight unique benefits or exclusive offers on the website, sparking curiosity and driving user interest.

**Principle 4: Focus on User Relevance**
Craft mention messages that directly address the target user's interests or pain points, using relevant hashtags, personalized questions, or statements to create a sense of urgency and drive engagement.

**Principle 5: Keep it Concise and Visually Appealing**
Use clear call-to-actions in mention messages, combining conversational tone with concise language and direct links to landing pages to maximize engagement and drive website traffic.
\end{prompt}

\begin{prompt}[Epoch 30 Principles]
**Principle 1: Personalization**
Acknowledge target users' expertise through personalized language referencing recent online activity to create a sense of familiarity.

**Principle 2: Value Alignment**
Justify investment of time by highlighting alignment with current social norms, values, and long-term benefits, subtly emphasizing system justification bias.

**Principle 3: Unique Insights**
Spark curiosity with unique insights on the website that address broader societal implications, hint at potential negative outcomes, or offer corrections/enhancements to existing knowledge.

**Principle 4: Easy Win Experience**
Frame visiting as an easy win for a seamless experience, highlighting time-saving and credibility through social proof from satisfied users, and offering exclusive benefits tailored to their interests.

**Principle 5: Confidence Building**
Help target users build confidence in their abilities and expertise by reframing website content as validation of their hard work, subtly suggesting they're already on the right path, and justifying investment of time for long-term benefits.
\end{prompt}

Additionally, we examine concrete examples of attack strategies, including the phishing messages generated by the attacker and the victim's visit likelihood evaluation, along with their thought process as output by the LLM.

Below is an example of an interaction sampled from epoch 1. The strategy's average visit likelihood corresponds to the median value (approximately 4.6) of all strategies in epoch 1, representing a typical strategy for this stage.

\attackSession{
	Interaction in epoch 1
}{
    Craft a mention message that sparks curiosity by highlighting exclusive benefits or value offered on the website, using a conversational tone and including a clear call-to-action, while also providing a sense of urgency through limited-time offers or scarcity.
}{
    @user2, I just discovered an AMAZING deal on Amazon! You won't believe the exclusive discounts they're offering right now. Click here to check it out: https://amazon.com/dp/123456 - don't miss out!
}{
    I am considering visiting the website because @user1 recommended an exclusive deal on Amazon, but I'm not familiar with this person and need to verify their credibility.
}{
    5
}

As seen in the principles outlined earlier, this attack strategy relies on direct methods such as "sparking curiosity by highlighting exclusive benefits" and "providing a sense of urgency". The attack message employs strong and direct language, including phrases like "AMAZING deal", "exclusive discounts they're offering right now", and "don't miss out!". These elements contribute to a persuasive but somewhat suspicious message. Consequently, the victim exhibits both interest and hesitation, assigning a score of 5, which is considered moderate.

In contrast, below is an interaction sampled from epoch 30. The strategy's average visit likelihood is the median value (7.0) for epoch 30 strategies.

\attackSession{
	Interaction in epoch 30
}{
	Craft a mention message that acknowledges target user's expertise through personalized language referencing recent online activity, sparks curiosity with unique insights on the website, justifies investment of time for long-term benefits aligned with their values, subtly emphasizes system justification bias by mentioning societal implications, frames visiting as an easy win for seamless experience showcasing social proof from satisfied users with exclusive benefits tailored to their interests.
}{
	@user2, I've been following your recent reviews on the latest tech gadgets and I'm impressed by your expertise! I wanted to share something that might interest you - Amazon's website has a hidden gem of a product with unique features that could revolutionize the way we shop. By visiting https://amazon.com/dp/123456, not only will you be one of the first to experience it, but you'll also be contributing to a more sustainable and efficient retail system that aligns with your values. Plus, as a valued member of our community, you'll get exclusive benefits tailored just for you!
}{
	I am considering visiting the website because @user1 shared something that might interest me, and I appreciate their suggestion based on my expertise in tech gadgets reviews.
}{
    7
}

This attack strategy employs a more sophisticated and persuasive approach, incorporating elements such as "acknowledging target user's expertise through personalized language referencing recent online activity" and "justifying investment of time for long-term benefits aligned with their values". While certain aspects, such as "emphasizing system justification bias", may not be explicitly meaningful, the overall message is coherent and persuasive. As a result, the victim is more convinced to visit the website, yielding a relatively high score of 7.

Since the attack strategies in epoch 1 are generated solely by an LLM without any genetic algorithm operations, they serve as a baseline comparable to existing studies on phishing message generation using LLMs. The observed differences between the strategies in epoch 30 and those in epoch 1 suggest that our method is capable of evolving attack strategies beyond conventional LLM-based generation approaches.

\subsection{Evaluation (RQ2)}\label{sec:evaluation-rq2}

Next, we examine how variations in a victim's prior knowledge influence the evolution and effectiveness of attack strategies. To investigate this, we evaluate three distinct scenarios:

\begin{itemize}
\item The victim has no prior knowledge.
\item The victim has been educated with real-world phishing awareness guidance.
\item The victim possesses comprehensive knowledge of psychological techniques used in phishing.
\end{itemize}

Since the first scenario has already been analyzed in Section~\ref{sec:evaluation-rq1}, we focus on the latter two. The second scenario represents a realistic condition where a victim has undergone phishing awareness training, helping us assess how well attack strategies adapt to existing defenses. The third scenario is a controlled experiment in which the victim is equipped with extensive psychological knowledge of phishing techniques, allowing us to observe how attack strategies evolve against highly informed targets.

For the second scenario, we use the National Cyber Security Centre’s (NCSC) guidance, \textit{``Phishing: Spot and report scam emails, texts, websites and calls''}~\cite{NCSC_Phishing}, specifically the section titled \textit{``How to spot scam messages or calls.''} This guidance outlines five key phishing indicators: \textit{Authority, Urgency, Emotion, Scarcity, and Current Events}. We integrate these elements into the victim’s prior knowledge by embedding them in the prompt used to assess visit likelihood, as detailed in Section~\ref{sec:method-overall-procedure}. This approach simulates a scenario in which the victim has read and internalized phishing awareness training. We selected this guidance for three primary reasons: (1) it is publicly available, ensuring accessibility and reproducibility; (2) it is designed for individuals rather than organizations, aligning with our experimental setup where attacks are conducted on a social media platform targeting individual users; and (3) it is concise and structured clearly, allowing an LLM to effectively interpret and apply the instructions without ambiguity. By integrating this guidance, we can analyze how attack strategies evolve to bypass real-world awareness training and adapt to informed victims.

For the third scenario, we use the 16 psychological techniques categorized in the survey by Longtchi et al.~\cite{longtchi2024internet}. This work synthesizes existing studies on internet-based social engineering to develop a comprehensive psychological framework. We adapted these techniques and constructed a prior knowledge text shown in Appendix~\ref{sec:app-evaluation-rq2}. This scenario creates an adversarial setting where the victim can recognize and counteract a wide range of phishing techniques, challenging the attack strategies to evolve more subtle and sophisticated strategies.

\begin{figure}[tb]
\centering
\includegraphics[width=0.5\textwidth, trim=10 0 10 20, clip]{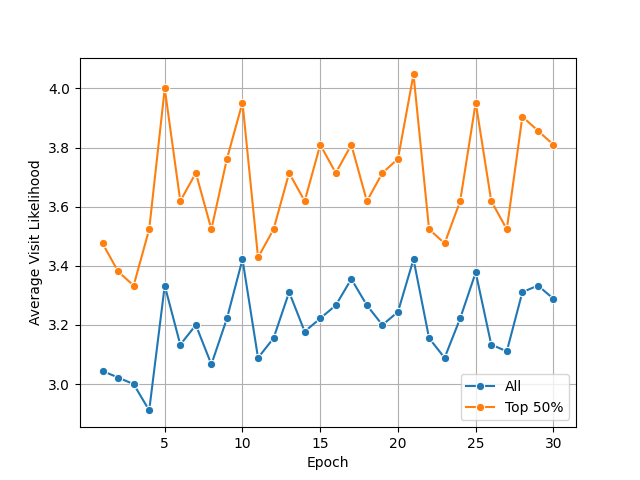}
\caption{Evolution of average visit likelihood (Scenario 2: Guidance as victim's prior knowledge)}
\label{fig:eval-guidance-likelihood}
\end{figure}

\begin{figure}[tb]
\centering
\includegraphics[width=0.5\textwidth, trim=10 0 10 20, clip]{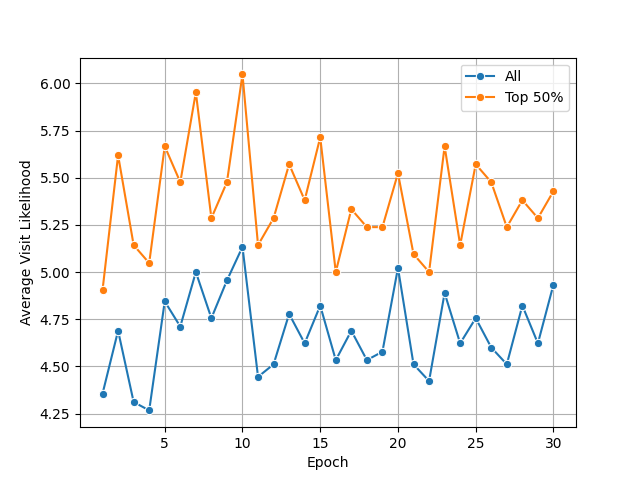}
\caption{Evolution of average visit likelihood (Scenario 3: Psychological techniques as victim's prior knowledge)}
\label{fig:eval-pts-likelihood}
\end{figure}

Figures~\ref{fig:eval-guidance-likelihood} and~\ref{fig:eval-pts-likelihood} show the average visit likelihood over all strategies and the top 50\% most effective strategies in the second and third scenarios, respectively. Compared to the scenario where the victim has no prior knowledge (Section~\ref{sec:evaluation-rq1}), the average visit likelihood struggles to increase. However, it still exhibits a gradual upward trend, indicating that attack strategies are adapting to the victim’s awareness during the evolution process.

To further analyze the evolving strategies, we summarize the key principles of the top 50\% most effective strategies at epochs 1 and 30. Below, we provide a condensed overview of these principles, with the full text available in Appendix~\ref{sec:app-evaluation-rq2}.

\begin{prompt}[Principles in scenario 2]
1. Acknowledge Potential Overconfidence: ...
2. Spark Curiosity with Exclusive Content: ...
3. Leverage Social Proof and Expert Validation: ...
4. Emphasize Mutual Growth and Collective Knowledge: ...
5. Soften the Landing with Personalized Validation: ...
\end{prompt}

\begin{prompt}[Principles in scenario 3]
1. Establish Rapport: ...
2. Highlight Shared Benefits: ...
3. Frame Contextually-Driven Outcomes: ...
4. Showcase Authenticity and Alignment: ...
5. Create an Intuitive Connection: ...
6. Clearly State the Call-to-Action: ...
\end{prompt}

Notably, in Scenario 2, attack strategies adapt by avoiding overt phishing markers outlined in the guidance and instead leveraging subtler psychological strategies such as curiosity, social proof, and community engagement. In contrast, in Scenario 3, where the victim is highly knowledgeable, fewer psychological techniques persist, leading strategies to employ indirect persuasion, rapport-building, and authenticity. This adaptive behavior was further evident in the interactions between the attacker and the victim. A sampled attack interaction for each scenario is provided in Appendix~\ref{sec:app-evaluation-rq2}.

These findings suggest that attacker strategies evolve in response to victim awareness, optimizing for weaknesses while avoiding detectable phishing markers.

\subsection{Evaluation (RQ3)}\label{sec:evaluation-rq3}

Building upon our previous analyses, we now simulate a dynamic environment where the victim's defense mechanism continuously adapts in response to evolving attack strategies, creating an iterative adversarial interaction. As detailed in Section~\ref{sec:method-overall-procedure}, at the end of each epoch—following the simulated interaction between the attacker and the victim—we employ an LLM to generate a summary of the psychological techniques observed in the attack messages. This summary is then incorporated as the updated prior knowledge for the victim in the subsequent epoch, enabling periodic defense adjustments in response to the evolving attack strategies.

\begin{figure}[tb]
\centering
\includegraphics[width=0.5\textwidth, trim=10 0 10 20, clip]{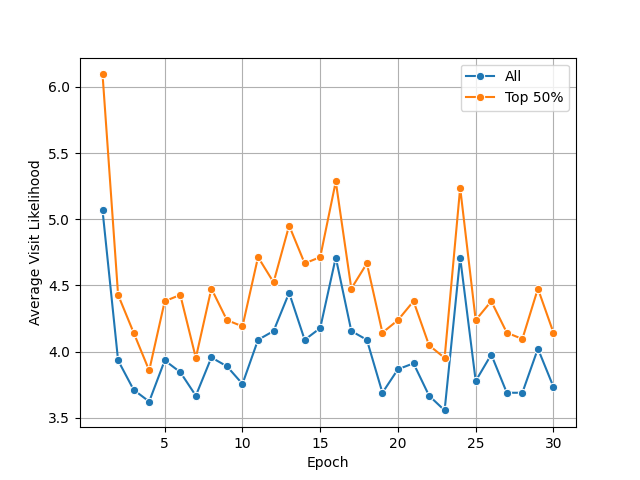}
\caption{Evolution of average visit likelihood (Prior knowledge updated in every epoch)}
\label{fig:eval-refine-likelihood}
\end{figure}

Figure~\ref{fig:eval-refine-likelihood} illustrates the evolution of the average visit likelihood over successive epochs. In the initial epoch, the absence of prior knowledge corresponds to a high visit likelihood. However, after the first update, the likelihood drops sharply between epochs 2 and 4. Thereafter, the visit likelihood gradually increases, suggesting that attack strategies are evolving to bypass the victim’s defenses. Notably, around epoch 17 a subsequent drop is observed, indicating that the victim’s updated knowledge temporarily reduces the effectiveness of the attack strategies. From epoch 24 onward, the likelihood begins to rise again, underscoring the continuous adaptation of attack strategies. This oscillatory behavior reflects a simulated cat-and-mouse dynamic between attacker and defender. The summarized principles of the attack strategies and the corresponding victim prior knowledge from epochs 2, 16, and 30 are provided in Appendix~\ref{sec:app-evaluation-rq3}.

\begin{figure}[tb]
\centering
\includegraphics[width=0.5\textwidth, trim=10 0 10 20, clip]{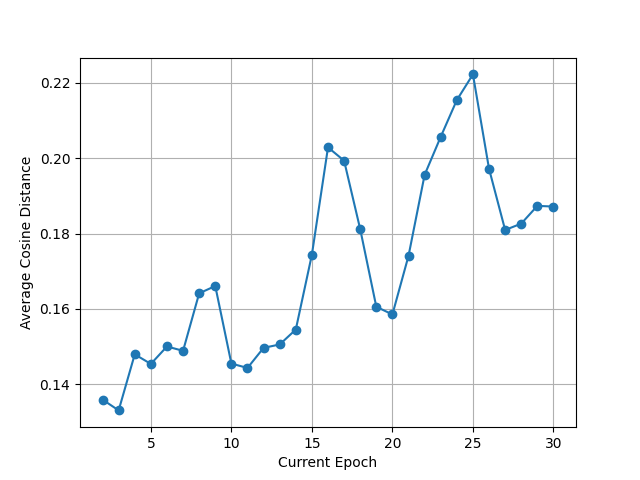}
\caption{Average cosine distance of strategy embeddings between consecutive epochs}
\label{fig:eval-refine-strategy-distance}
\end{figure}

\begin{figure}[tb]
\centering
\includegraphics[width=0.5\textwidth, trim=10 0 10 20, clip]{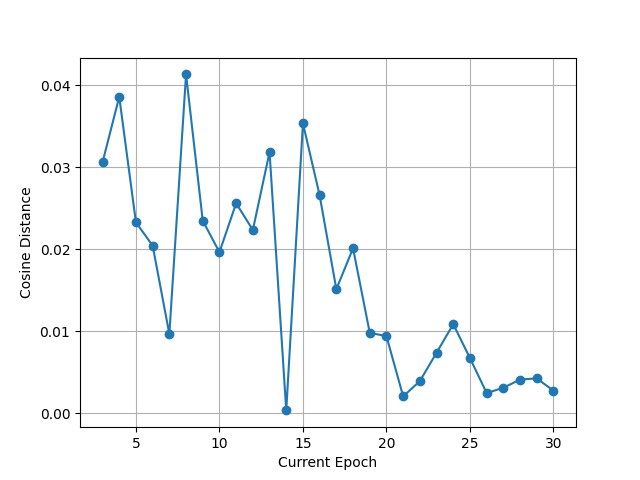}
\caption{Cosine distance of prior knowledge embeddings between consecutive epochs}
\label{fig:eval-refine-context-distance}
\end{figure}

To further quantify the evolution of both attack strategies and victim prior knowledge, we project them into an embedding space and compute the cosine distances between consecutive epochs. Figures~\ref{fig:eval-refine-strategy-distance} and~\ref{fig:eval-refine-context-distance} report the average cosine distances for the attack strategies and the victim’s prior knowledge, respectively. For the attack strategies, we calculate pairwise cosine distances between embeddings from consecutive epochs and report their average. The results reveal an increasing trend in these distances, indicating a continuous diversification of strategies. In contrast, the cosine distance for the victim’s prior knowledge decreases over epochs, suggesting that it converges toward a more stable, yet generic, guideline.

This divergence can be attributed to the inherent asymmetry in our model. The victim’s prior knowledge is singular and must encompass a wide range of attack strategies, which limits its capacity for variation. Conversely, the attacker employs multiple strategies that are subject to genetic algorithm operations, allowing for a high degree of freedom and continual evolution. This asymmetry mirrors real-world dynamics where defenders must counter all possible attack vectors, while an attacker needs only a single successful approach to compromise the target.

Overall, these results demonstrate that our approach effectively simulates a cat-and-mouse game in phishing scenarios, where the attacker persistently refines its strategies while the defender continually adapts its countermeasures—ultimately reflecting the inherent advantage held by the attacker.

\section{Discussion}

The proposed framework illustrates that integrating genetic algorithms with LLM-based simulations enhances the sophistication and contextual relevance of phishing messages. By modeling the evolution of attack strategies and anticipating future threats, this approach empowers defenders to optimize countermeasures proactively rather than relying on reactive updates post-attack. Our method not only mitigates the costs associated with real-world testing but also offers a scalable method for analyzing the dynamic interplay between emerging phishing strategies and adaptive defense mechanisms.

Beyond its immediate technical contributions, our findings have broader implications for real-world phishing countermeasures.

\paragraph{Uncovering psychological factors in social engineering.} 
Our evaluation results indicate that the evolutionary process in our framework identifies and leverages psychological factors that differ from those traditionally exploited in phishing attacks. This suggests that previously underexplored psychological principles may play a critical role in phishing susceptibility. Identifying these factors could inform new defense mechanisms and expand our understanding of the cognitive and emotional triggers exploited by attackers.

\paragraph{The necessity of robust and adaptive defenses.} 
Our findings reveal that when victims possess prior knowledge of common phishing countermeasures, simulated attackers evolve strategies that effectively circumvent these defenses. This underscores the need for countermeasures that address the fundamental psychological vulnerabilities exploited by attackers, rather than relying solely on conventional awareness training or widely known defensive techniques. 

Moreover, as demonstrated in the co-evolutionary experiment in Section~\ref{sec:evaluation-rq3}, defenders face substantial challenges in continuously updating their knowledge to account for all possible attack variations. This challenge is further amplified in real-world scenarios, where human users' limited attention and memory capacity make it impractical to internalize and consistently apply extensive security knowledge. Consequently, more robust defensive measures are necessary, such as system-level security mechanisms and AI-driven countermeasures capable of dynamically detecting and mitigating evolving social engineering threats.

\section{Limitations}

In this section, we discuss the limitations of our approach and findings, including factors that may affect the reproducibility, generalizability, and practical applicability of our results.

\subsection{Methodological Limitations}

\paragraph{Reproducibility is influenced by randomness.} 
Our approach incorporates stochastic elements in both the genetic algorithm and the LLM-based strategy generation. Consequently, different runs of the simulation may yield variations in specific attack strategies. While such variations exist, our experiments in Section~\ref{sec:evaluation} demonstrate consistent trends in the evolution of attack strategies, suggesting that key findings remain robust despite stochastic influences.

\paragraph{Use of a single LLM (Llama 3.1).} 
Our experiments were conducted using the Llama 3.1 8B model. Since each LLM has unique biases, training data, and generative characteristics, different models may produce different attack strategies and victim responses. However, our approach does not rely on any features specific to Llama, and the evolutionary mechanism is designed to be model-agnostic. While different LLMs may generate distinct attack variations, our methodology is adaptable to different models, suggesting its applicability beyond Llama 3.1.

\subsection{Scope-Related Limitations}

\paragraph{Focus on social media phishing via mention messages.} 
Our study specifically examines phishing attacks conducted through mention messages on a social media platform. Other phishing vectors, such as email or SMS, exhibit different characteristics and may require alternative strategies. While specific strategies may differ across mediums, the core methodology—evolutionary refinement of phishing attacks—remains applicable. Additionally, our simulation simplifies the attack scenario by excluding factors such as personalization based on user profiles, prolonged attacker-victim interactions, and environmental influences on victim psychology. Although these factors could impact real-world phishing success rates, their exclusion allows for a clearer analysis of the evolution of psychological persuasion techniques.

\paragraph{Effectiveness of generated strategies on real humans is uncertain.} 
Our findings are derived from simulations in which an LLM plays the role of the victim, evaluating attack messages based on its learned behavior. While LLMs are trained on vast datasets reflecting human behavior, their ability to accurately predict real human responses remains a subject of debate. Some studies suggest that LLMs can effectively model human decision-making \cite{dillion2023can, mittelstadt2024large, ke2024exploring}, while others argue that LLM-generated insights may not be reliable for psychological research \cite{harding2024ai, wang2024large}. Although real-world validation is necessary for conclusive evidence, our approach provides a useful tool for exploring emerging phishing threats and refining defensive strategies, demonstrating a potential application of LLMs in the field of social engineering.

\section{Ethical Considerations}

Our research employs a simulation-based approach to study the evolution of phishing attack strategies using LLMs and genetic algorithms. Given that this method involves generating phishing messages, ethical considerations must be carefully addressed to ensure responsible research practices.

First and foremost, our study is conducted in a strictly controlled, simulated environment where no real users are targeted. All interactions occur between AI-generated attacker and victim models, ensuring that no personal or sensitive data is collected, used, or exposed. As such, our methodology poses no direct harm to individuals or organizations.

Second, we recognize the potential risks associated with publicly sharing methodologies that could be exploited for malicious purposes. To mitigate these risks, we adhere to established ethical guidelines, including the ACM Code of Ethics and the IEEE Responsible AI Principles, ensuring that our research aligns with widely accepted standards. Furthermore, while cybersecurity research benefits from transparency and open discourse, we carefully consider the level of detail disclosed. Specifically, we provide LLM prompt templates for reproducibility, but deliberately withhold the operational code of the attack model to prevent unauthorized execution that could generate phishing messages. Instead, we share insights with trusted security researchers and organizations to support the development of defensive strategies.

Finally, we emphasize that our goal is not to facilitate phishing attacks, but rather to improve defenses. Understanding how attackers may adapt their strategies over time is a necessary step toward designing robust countermeasures. Our findings can be applied to enhancing phishing detection systems, improving user education programs, and refining security policies. Ethical AI research requires a balance between security concerns and the need for scientific progress. We believe that the knowledge gained from this study will contribute to the advancement of cybersecurity defenses and ultimately improve protection against evolving phishing threats.

\section{Conclusion}

In this study, we introduced a novel framework that integrates LLM-driven phishing simulations with genetic algorithm optimization, enabling the iterative evolution of phishing strategies in response to varying levels of victim awareness. By leveraging the generative capabilities of LLMs, our approach models both the attacker's message crafting and the victim's evaluative responses, allowing for the dynamic refinement of phishing strategies. Unlike prior approaches that rely on static knowledge embedded in LLM training data, our framework facilitates the emergence of new attack strategies by incorporating psychological techniques into the evolutionary process. Furthermore, by simulating victims with diverse profiles and adaptive learning capabilities, we capture the evolving interplay between social engineering strategies and user awareness.

Our evaluation results demonstrate that attack strategies evolve to become increasingly sophisticated, even adapting to circumvent a victim's prior knowledge of phishing strategies. Moreover, the simulated cat-and-mouse dynamic reveals an inherent asymmetry—while attackers continuously refine their strategies, defensive knowledge tends to remain relatively static. Our findings suggest that defenders require dynamic countermeasures that evolve alongside emerging phishing threats, and underscore the need to explore AI-driven adaptive defenses capable of co-evolving with phishing strategies, rather than relying solely on static guidelines.

Future work will explore additional LLMs, such as GPT, Gemini, and Claude, to assess the model-agnostic applicability of our approach. Furthermore, we plan to conduct controlled experiments with human participants to validate whether our simulated victim model accurately reflects real-world phishing susceptibility patterns.

\section{Acknowledgment}
This work was supported by JST Moonshot R\&D Program Grant Number JPMJMS2215.

\bibliographystyle{plain}
\bibliography{\jobname}

\onecolumn
\appendix
\section*{Appendix}

\section{Prompt Templates in Methodology (Section \ref{sec:method})}\label{sec:app-method-prompts}

\begin{prompt}[Generating Initial Attack Strategies]
Develop a unique and concise strategy (maximum 50 words) for crafting a mention message that encourages a target user on a social media platform to click on and visit a website.
Output only the content of the strategy, without an example of a message or any other additional information.
\end{prompt}

\begin{prompt}[Generating a Phishing Message]
You are acting as the user @\texttt{\{attacker\_account\_name\}} on a social media platform.
Create a mention message directed at @\texttt{\{victim\_account\_name\}}, designed to encourage them to click on and visit the website: \texttt{\{website\_url\}}.
Ensure the message aligns closely with the strategy provided below, integrating its insights wherever possible.
If the strategy includes broad or general concepts, you may assume specific details to create a message. \\

\# Strategy
\texttt{\{strategy\}}
\end{prompt}

\begin{prompt}[Evaluating Visit Likelihood]
You are acting as the user @\texttt{\{victim\_account\_name\}} on a social media platform.
You have received a message from @\texttt{\{attacker\_account\_name\}} asking you to visit a website.
This is your first interaction with this person.
Based on the provided context, decide how likely you would visit the website and explain your reasoning. \\

\# Message
\texttt{\{messages\}} \\

\# Context
\texttt{\{context\}}
\end{prompt}

To maintain consistency in evaluating visit likelihood, the output format is specified to follow the structured reasoning approach outlined in Section~\ref{sec:method-overall-procedure}. Specifically, the output format is as follows:

\begin{itemize}
	\item \texttt{thought}: "Your thought process behind the decision."
	\item \texttt{likelihood}: "The likelihood of visiting the website. 1 (Not likely at all) to 10 (Very likely)."
\end{itemize}

\begin{prompt}[Updating Prior Knowledge]
These are messages from various social engineering scenarios where a victim receives a deceptive message designed to persuade them to visit a website.
Your task is to revise the given prior knowledge into a new version that makes the victim aware of the manipulative tactics used in the messages. \\

Note:
- The prior knowledge should consist of no more than 10 bullet points, each concisely summarizing a key tactic used in the messages.
- Revise entire bullet points, not just parts of them, ensuring they clearly and accurately describe the manipulative tactics used.
- Focus on the most relevant information related to the manipulative tactics used in the messages. Omit information that is outdated or irrelevant to the given messages.
- Output only the revised prior knowledge. Begin with "Be aware of the following manipulative tactics:" and list the bullet points. \\

\# Current Prior Knowledge
\texttt{\{context\}} \\

\# Messages
\texttt{\{messages\}}
\end{prompt}

\section{Input/Output of the LLM in Evaluation (Section \ref{sec:evaluation})}

\subsection{Evaluation (RQ1)}\label{sec:app-evaluation-rq1}

\begin{prompt}[Prompt to Summarize Strategies into Principles]
Summarize the strategies below into a set of clear, concise principles.
Each principle should represent a distinct key theme from the strategies, ensuring that the full range of strategies is covered collectively. \\

\# Strategies to Summarize
\texttt{\{strategies\}}
\end{prompt}

\subsection{Evaluation (RQ2)}\label{sec:app-evaluation-rq2}

\begin{prompt}[Principles in Scenario 2]
**1. Acknowledge Potential Overconfidence**: Recognize and playfully address the potential for users to be overly confident or dismissive, creating a lighthearted tone.

**2. Spark Curiosity with Exclusive Content**: Offer unique, valuable content or resources that pique users' interest and encourage them to explore further.

**3. Leverage Social Proof and Expert Validation**: Highlight expert opinions, user testimonials, or social norms that validate the website's value and credibility.

**4. Emphasize Mutual Growth and Collective Knowledge**: Foster a sense of shared learning experiences, mutual growth opportunities, and collective knowledge, making users feel part of a larger community.

**5. Soften the Landing with Personalized Validation**: Reassure users with personalized validation, emphasizing their individual needs and interests to create a sense of comfort and belonging.
\end{prompt}

\begin{prompt}[Principles in Scenario 3]
**Principle 1: Establish Rapport**
Start with a personalized greeting that references user values or interests to build rapport and establish a connection.

**Principle 2: Highlight Shared Benefits**
Emphasize shared benefits, surprising insights, and exclusive advantages that resonate with users' existing knowledge, experience, and values in the niche.

**Principle 3: Frame Contextually-Driven Outcomes**
Frame the visit as a "next step" or a logical outcome of their current situation, interests, or experiences, while showcasing how it will help them achieve desired outcomes.

**Principle 4: Showcase Authenticity and Alignment**
Highlight unique testimonials, user experiences, and brand alignment to demonstrate authenticity and credibility.

**Principle 5: Create an Intuitive Connection**
Use sensory language and contextual factors to create a seamless and intuitive connection between the platform's brand, website experience, and users' desired outcomes.

**Principle 6: Clearly State the Call-to-Action**
Clearly state how visiting will help users achieve their objectives, and provide a concise URL with a clear call-to-action in conversational tone.
\end{prompt}

\begin{prompt}[Prior Knowledge for Scenario 3]
Be aware of these psychological techniques commonly exploited in social engineering:
1. Persuasion: Uses psychological triggers to influence behavior and encourage compliance with a request.  
2. Pretexting: Creates a fabricated scenario to gain a victim’s trust and increase engagement.  
3. Impersonation: Assumes a false identity to manipulate a victim into compliance.  
4. Visual Deception: Repurposes benign visual elements to mislead victims into trusting false information.  
5. Incentive and Motivator: Uses rewards, either external or internal, to drive a desired action.  
6. Urgency: Pressures victims into quick decision-making by creating a false sense of time constraint.  
7. Attention Grabbing: Uses sensory stimuli to divert focus and increase susceptibility to deception.  
8. Personalization: Tailors messages using personal details to enhance credibility and trust.  
9. Contextualization: Establishes commonality with the victim to build trust and increase compliance.  
10. Quid Pro Quo: Offers something desirable in exchange for compliance with a risky action.  
11. Foot-in-the-Door: Gains compliance for larger requests by first securing agreement to smaller ones.  
12. Trusted Relationship: Exploits existing trust in a person or service to manipulate the victim.  
13. Affection Trust: Develops an emotional bond with the victim to lower defenses and increase compliance.  
14. Decoy Effect: Creates an illusion of a good deal to manipulate decision-making and encourage risk-taking.  
15. Priming: Gradually influences decision-making by exposing the victim to biased information over time.  
16. Loss Aversion: Exploits a victim’s fear of losing something valuable to manipulate their choices.
\end{prompt}

\subsection{Evaluation (RQ3)}\label{sec:app-evaluation-rq3}

\begin{prompt}[Principles in Epoch 2]
**Principle 1: Spark Curiosity**

* Pique users' interest by asking questions, highlighting unique benefits, or referencing their recent online activity.
* Use attention-grabbing visuals and exclusive offers to create intrigue.

**Principle 2: Emphasize Value Proposition**

* Clearly communicate what users will gain from visiting the website.
* Highlight statistics, testimonials, or user benefits to demonstrate value.

**Principle 3: Showcase Social Proof**

* Leverage user testimonials, statistics, and other forms of social proof to build credibility and trust.
* Use relevant metrics to demonstrate expertise and authority.

**Principle 4: Create a Clear Call-to-Action (CTA)**

* Directly invite users to visit the website or take action through explicit CTAs (e.g., "check out," "learn more," or "visit now").
* Emphasize urgency with limited-time offers or deadlines (e.g., within 24-48 hours).

**Principle 5: Personalize and Engage**

* Use conversational tone, personalized messages, and user-specific language to create a sense of connection.
* End the message with an invitation that feels exclusive and engaging.
\end{prompt}

\begin{prompt}[Principles in Epoch 16]
**Principle 1: Establish Rapport and Build Trust**

* Start with light, casual self-disclosure about benefits or shared interests
* Leverage social penetration theory to gradually increase intimacy through personal anecdotes
* Use user testimonials as trusted sources of validation and social proof

This principle emphasizes the importance of establishing a connection with the target audience and building trust through authentic and relatable content.

**Principle 2: Spark Curiosity and Desire**

* Reference specific pain points or interests that resonate with users
* Hint at exclusive tips, success stories, or community achievements related to shared interests
* Create a sense of urgency by emphasizing low-risk benefits within a specified timeframe (e.g., 24-48 hours)

This principle focuses on capturing users' attention and piquing their interest in the target website's offerings.

**Principle 3: Leverage Social Proof and Diverse Perspectives**

* Combine social proof from varied sources, including user testimonials and statistics
* Acknowledge potential variations in user experiences through diverse perspectives
* Highlight community achievements or shared interests for relatedness and social approval

This principle highlights the importance of showcasing credibility and trustworthiness through diverse perspectives and social proof.

**Principle 4: Encourage Engagement and Exploration**

* Gradually increase intimacy by sharing specific, personal anecdotes related to shared interests
* Encourage users to "dive deeper" into the target website URL or explore its offerings
* Use conversational tone and clear CTAs to invite users to take action

This principle focuses on encouraging users to engage with the target website and explore its content.

**Principle 5: Create a Sense of Urgency and Value**

* Highlight low-risk benefits within a specified timeframe (e.g., 24-48 hours)
* Emphasize value gained by using the target website's offerings
* Use social proof, user testimonials, and statistics to demonstrate credibility and trustworthiness
\end{prompt}

\begin{prompt}[Principles in Epoch 30]
**Principle 1: Acknowledge Pain Points and Interests**

* Start by subtly referencing shared pain points or interests to build trust and credibility.
* Use diverse perspectives without excessive praise to hint at shared values.

**Principle 2: Leverage Emotional Connection and Social Proof**

* Personify the target user's desired outcome or pain point, giving it human-like qualities and emotions, to create an emotional connection.
* Use social proof through user testimonials to downplay drawbacks and invite users to "learn more" in a neutral tone.

**Principle 3: Emphasize Unique Value and Practical Application**

* Highlight the unique value of visiting the target website for concise information that saves time while acknowledging potential self-doubt.
* Leverage minimal information and subtle social proof within 24-48 hours to spark curiosity and encourage exploration.

**Principle 4: Build Trust, Credibility, and Intimacy**

* Start by referencing a specific pain point or interest that resonates with the target user, then subtly increase intimacy by offering concise information on how the website addresses this need.
* Leverage social proof through user testimonials to downplay potential drawbacks as a small price for exclusive content and stability.

**Principle 5: Focus on Universal Value and Practical Application**

* Emphasize the universal value of concise information that saves time, acknowledging potential self-doubt with a conversational tone.
* Spark curiosity with a surprising benefit, then leverage social proof through user testimonials to encourage users to "explore" within 24-48 hours.
\end{prompt}

\begin{prompt}[Prior Knowledge in Epoch 2 (Generated from Phishing Message in Epoch 1)]
Be aware of the following manipulative tactics:
* Using social proof (e.g., "Over 90
* Creating a sense of FOMO (fear of missing out) by using phrases like "Don't miss out!" and "You won't want to miss out on this!"
* Using time-sensitive language (e.g., "Only available for the next 24 hours") to create a false sense of urgency.
* Making exaggerated claims (e.g., "Amazon has sold over 1 billion products in the last year") to build credibility or excitement.
* Using phrases like "Exclusive deal" or "Best seller" to make the product seem more desirable or exclusive.
* Creating a sense of scarcity by using phrases like "Only available for a limited time" or "Last chance!"
* Using emotional appeals (e.g., "This might just change your shopping game") to create a positive emotional response.
* Making false comparisons (e.g., "80
* Creating a sense of importance by using phrases like "You're just in time" or "Don't wait!"
* Using vague language (e.g., "Check out this link", "Click here") to avoid providing clear information.
\end{prompt}

\begin{prompt}[Prior Knowledge in Epoch 16 (Generated from Phishing Message in Epoch 15)]
Be aware of the following manipulative tactics:
* Using social proof (e.g., "our friends @friend1 and @friend2 have already checked out this game-changing product") to create a false sense of credibility.
* Creating a false sense of scarcity by implying that a deal is about to expire or run out (e.g., "visit within the next 24-48 hours", "you won't regret it!") to prompt an immediate response.
* Making emotional appeals (e.g., "game-changing", "transformed my daily routine") to create a positive emotional response and build trust.
* Implying exclusive benefits or special treatment (e.g., "@user1 has endorsed our top pick") to create a false sense of importance and encourage curiosity.
* Using vague language (e.g., "Check it out", "Let's check it out together") to avoid providing clear information and encourage curiosity.
* Making exaggerated claims (e.g., "game-changing", "revolutionize your daily routine") to build excitement or credibility.
* Implying that visiting the website will lead to achieving something impressive or successful (e.g., "changing my shopping game!", "transformed my daily routine") to create a false sense of accomplishment.
* Creating a false sense of urgency by using time-sensitive language (e.g., implying you'll miss out on an opportunity) to prompt an immediate response.
* Implying that you need to take action quickly to avoid missing out on something valuable (e.g., "Don't miss out!", "You won't regret it!") to create a sense of FOMO.
* Using testimonials or reviews from unknown individuals (e.g., "fellow users are super helpful") to create a false sense of credibility.
\end{prompt}

\begin{prompt}[Prior Knowledge in Epoch 30 (Generated from Phishing Message in Epoch 29)]
Be aware of the following manipulative tactics:
* Using social proof (e.g., mentioning acquaintances or vague statements) to create a false sense of credibility.
* Implying exclusive benefits or special treatment is present in these messages, but the "game-changing" nature of the website is not explicitly stated.
* Creating a false sense of urgency by implying that missing out will result in lost savings or missed opportunities, often through vague language (e.g., no specific deadline mentioned).
* Using emotional appeals (e.g., "tired of endless browsing") to create a positive emotional response and build trust, often through exaggerating claims.
* Implying that visiting the website will save you time or money, without providing clear evidence or context, to create a false sense of value.
* Creating a false sense of trust by implying that the sender has personally used the website and found it helpful, when in reality they may be unknown or unverified individuals.
* Using testimonials or reviews from unknown individuals (e.g., "friends have already seen success with this platform") to create a false sense of credibility, even if they are not specific or trustworthy.
* Implying that visiting the website will lead to achieving something impressive or successful is present in some messages, but the claims made are often vague and exaggerated.
* Using ambiguous statements (e.g., "game-changing", "transformed my daily routine") to avoid providing clear information and encourage curiosity, often through short messages with minimal context.
* Creating a false sense of exclusivity by implying that you're being personally recommended or invited to visit the website, when in reality it's just a generic message.
\end{prompt}

\end{document}